\documentstyle[12pt]{article}
\newcommand{\be}{\begin{equation}}
\newcommand{\ee}{\end{equation}}
\newcommand{\ba}{\begin{eqnarray}}
\newcommand{\ea}{\end{eqnarray}}

\def\gl#1{(\ref{#1})}

\begin{document}

\begin{center}
{\Large\bf Vector mesons in Quasilocal Quark Models}\footnote{Talk given at
the XV International Workshop on High Energy Physics and
Quantum Field Theory, Tver',Russia, Sept.14-20,2000}\\

\vspace{4mm}

A. A. Andrianov\footnote{Visiting professor at the Departament d'Estructura
i Constituents de la Mat\`eria, Universitat de
Barcelona, Espa\~na.}, V. A. Andrianov and S. S. Afonin \\
Department of Theoretical Physics, St.Petersburg State
University,\\
198504 St.Petersburg, Russia \\
\end{center} 

\begin{abstract}
We consider the Quasilocal Quark Model of NJL type  
including vector and
axial-vector four-fermion interaction with derivatives. In the presence of
 a strong attraction in the scalar channel the chiral symmetry 
is spontaneously broken 
and as a consequence the composite meson states are  generated in the
vector and axial-vector channels. The
appropriate set of Chiral Symmetry Restoration Sum Rules in these 
channels are imposed as matching rules to QCD at
intermediate energies. The mass spectrum and some decay constants for ground
and excited  meson states
are calculated.
\end{abstract}

\bigskip

{\large\bf 1. Introduction}\\

The QCD-inspired quark models with four-fermion
interaction  are often applied for
the effective description of low-energy QCD in the hadronization regime. The
local four-fermion interaction is involved to induce the dynamical
chiral symmetry breaking (DCSB) due to strong attraction in the
scalar channel. As a consequence, the dynamical quark mass
$m_{dyn}$ is created, as well as an isospin multiplet of pions, 
massless in the chiral limit,
 and a massive scalar meson with mass
$m_{\sigma}=2m_{dyn}$ arise. However it is known from
experiment that there are series of meson states
with equal quantum numbers and heavier masses, in particular,
$0^{-+}(\pi,\pi',\pi'',...)$; $0^{++}(\sigma
(f_{0}),\sigma',\sigma'',...)$; $1^{--}(\rho,\rho',\rho'',...)$.
Due to confinement, one expects an infinite number of such excited
states with increasing masses. Therefore in order to describe the physics of
those resonances at  intermediate energies one can extend the quark model
with local interaction of the Nambu-Jona-Lasinio (NJL) type [1-10]
taking into account  higher-dimensional quark operators 
with derivatives, i.e. quasilocal quark interactions [11]. For 
sufficiently strong couplings the new
operators promote the formation of additional meson states. 
Such a
quasilocal approach (see also [12-16]) represents a
systematic extension of the NJL-model towards the complete
effective action of QCD where many-fermion vertices with
derivatives possess the manifest chiral symmetry of
interaction, motivated by the soft momentum expansion of the
perturbative QCD effective action. 
In the effective action of QuasilocalModels of NJL type (QNJLM) 
the low-energy
gluon effects are  hidden in the coupling constants. 
The alternative schemes including the condensates of low-energy
gluons can be found in [6,17].

At intermediate energies the correlators of QNJLM can be matched
\cite{zap} to
the O.P.E. of QCD correlators \cite{[18]}. This matching realizes the 
correspondence to QCD and improves the predictivity of QNJLM.
It is based on
the large-$N_{c}$ approach which is equivalent to planar QCD
\cite{[16],[17]}. In this approximation 
the correlators of color-singlet quark currents are saturated
by narrow meson resonances. In particular, the two-point
correlators of scalar, pseudoscalar, vector and axial-vector
quark currents are represented by the sum of related meson poles
in Euclidean space-time:
\be
\Pi^{C}(p^{2})=\int d^{4}x\exp(ipx)\langle T(\bar{q}\Gamma
q(x)\bar{q}\Gamma q(0))
\rangle_{planar}=\sum_{n}\frac{Z_{n}^{C}}{p^{2}+m_{C,n}^{2}}
+D_{0}^{C}+D_{1}^{C}p^{2}, \label{cor1}
\ee
$$
C\equiv S,P,V,A; \qquad
\Gamma=i,\gamma_{5},\gamma_{\mu},\gamma_{\mu}\gamma_{5}; \qquad
D_{0},D_{1}=const.
$$
The last two terms both in the scalar-pseudoscalar and in the
vector-axial-vector channel represent a perturbative contribution
with $D_{0}$ and $D_{1}$ being contact terms required for the
regularization of infinite sums. On the other hand 
the high-energy asymptotics is
provided \cite{[18]} by the perturbation theory and the operator
product expansion due to asymptotic freedom of QCD. Therefrom the
above correlators increase at large $p^{2}$,
\be
\Pi^{C}(p^{2})\mid_{p^{2}\rightarrow\infty}\sim
p^{2} \ln\frac{p^{2}}{\mu^{2}}.
\ee
When comparing the two expressions one concludes that the
infinite series of resonances with the same quantum numbers
should exist in order to reproduce the perturbative asymptotics.

Meantime the differences of correlators of opposite-parity currents
rapidly decrease at large momenta \cite{zap,[25]}:
\be
(\Pi^{P}(p^{2})-\Pi^{S}(p^{2}))_{p^{2}\rightarrow\infty}\equiv
\frac{\Delta_{SP}}{p^{4}}+O(\frac{1}{p^{6}}), \qquad \Delta_{SP}\simeq
24\pi\alpha_{s}<\bar{q}q>^{2} \label{SP}
\ee
and \cite{[18],[20],[21]}
\be
(\Pi^{V}(p^{2})-\Pi^{A}(p^{2}))_{p^{2}\rightarrow\infty}\equiv
\frac{\Delta_{VA}}{p^{6}}+O(\frac{1}{p^{8}}), \qquad \Delta_{VA}\simeq
-16\pi\alpha_{s}<\bar{q}q>^{2} \label{VA}
\ee
where we have defined in the V,A channels
\be
\Pi_{\mu\nu}^{V,A}(p^{2})\equiv(-\delta_{\mu\nu}p^{2}+p_{\mu}p_{\nu})
\Pi^{V,A}(p^{2}),
\ee
and the vacuum dominance hypothesis \cite{[18]} in the
large-$N_{c}$ limit is adopted.

Therefore the chiral symmetry is restored at high energies and
the two above differences represent a genuine order parameter of
CSB in QCD. As they decrease rapidly at large momenta
one can perform the matching of QCD asymptotics 
by means of few lowest lying resonances that gives a number of
constraints from Chiral Symmetry restoration.
They may be used
both for obtaining some additional bounds on the model parameters 
and for calculating of some decay constants (see, some examples in 
\cite{[20],[23]} and
references therein). 
In the present work the QNJL model is considered with two channels where two 
pairs of vector and axial-vector mesons are
generated. Respectively it is expected to reproduce the lower
part of QCD meson spectrum in the planar limit and the leading asymptotics of
chiral symmetry restoration for higher energies. 

\bigskip

{\large\bf 2. Quasilocal Quark Model of NJL-type}\\

The low energy effective action of QCD in the quark
sector has a qualitatively different structure depending on
whether it is built at the CSB scale $\Lambda$ by means of perturbation
theory \cite{[11],[24]} or below the CSB scale when the formation of 
light pseudoscalar meson is
implemented manifestly \cite{[25],[23],[26]}. We will discuss only
the first case where the QNJL model \cite{[11]}
extends the NJL one with chiral symmetry broken due to strong
attractive four-fermion forces in the color-singlet scalar
channel. When dynamical quark masses
are supposed to be sufficiently smaller than the CSB scale (cutoff) the
minimal $n$-channel lagrangian of the QNJLM has \cite{[11],[14]} 
the following form,
\be
L=\bar{q}i\hat{\partial}q+L_{SP}^{I},
\ee
where indices $S,P$ denote the scalar and pseudoscalar case and
\be
L_{SP}^{I}=\frac{1}{4N_{f}N_{c}\Lambda^{2}}\sum_{k,l=1}^{n}a_{kl}
[\bar{q}f_{k}q\cdot\bar{q}f_{l}q+
\bar{q}f_{k}i\gamma_{5}q\cdot\bar{q}f_{l}i\gamma_{5}q].
\ee
$a_{kl}$ represents here a symmetric matrix of real coupling
constants. We will restrict ourselves by the case $n=2$ and
describe the ground meson states and their first excitations only.
Let us choose the polynomial formfactors  as being orthogonal on
the unit interval,
\be
\int_{0}^{1}f_{k}(\tau)f_{l}(\tau)d\tau=\delta_{kl}.
\ee
We select out here:
\be
f_{1}(\tau)=2-3\tau; \qquad f_{2}(\tau)=-\sqrt{3}\tau; \qquad
\tau\equiv-\frac{\partial^{2}}{\Lambda^{2}}.
\ee
As this model interpolates the low-energy QCD action
it is supplied with a cutoff $\Lambda$ (of order of the 
CSB scale) for virtual quark
momenta in quark loops. For simplicity we  neglect 
the isospin effects encoded in  quarks of different flavor,
$N_f =1$. Moreover the chiral limit $m_q = 0$ is implied throughout
this paper. 

Let us now incorporate the vector and axial-vector interaction
with appropriate matrix of couplings $b_{kl}$,
$$
L_{SPVA}^{I}=\frac{1}{4N_{c}\Lambda^{2}}\sum_{k,l=1}^{2}
\{a_{kl}[\bar{q}f_{k}q\cdot\bar{q}f_{l}q+
\bar{q}f_{k}i\gamma_{5}q\cdot\bar{q}f_{l}i\gamma_{5}q]
$$
\be
+b_{kl}[\bar{q}f_{k}i\gamma_{\mu}q\cdot\bar{q}f_{l}i\gamma_{\mu}q
+\bar{q}f_{k}i\gamma_{\mu}\gamma_{5}q\cdot\bar{q}f_{l}i\gamma_{\mu}\gamma_{5}q]\}.
\ee
Following
the standard procedure we introduce auxiliary scalar,
pseudoscalar, vector and axial-vector fields,
$$
L_{aux}^{I}=\sum_{k=1}^{2}i\bar{q}(\sigma_
{k}+i\gamma_{5}\pi_{k}+i\rho_{k,\mu}\gamma_{\mu}+ia_{k,\mu}\gamma_{\mu}\gamma_{5})f_{k}q
$$
\be
+N_{c}\Lambda^{2}\sum_{k,l=1}^{2}\{\sigma_{k}a_{kl}^{-1}\sigma_{l}+\pi_{k}a_{kl}^{-1}\pi_{l}
+\rho_{k,\mu}b_{kl}^{-1}\rho_{l,\mu}+a_{k,\mu}b_{kl}^{-1}a_{l,\mu}\}.
\ee
The observables should not depend on
the cutoff $\Lambda$. The  scale invariance is achieved by
appropriate prescription of cutoff dependence for effective
coupling constants $a_{kl},b_{kl}$. Namely, we require the
cancellation of quadratic divergences and
parametrize the matrices of coupling constants in the vicinity
of polycritical point as follows,
$$
8\pi^{2}a_{kl}^{-1}=\delta_{kl}-\frac{\Delta_{kl}}{\Lambda^{2}};
\qquad \Delta_{kl}\ll\Lambda^{2}
$$
\be
16\pi^{2}b_{kl}^{-1}=\delta_{kl}-\frac{4}{3}
\frac{\bar\Delta_{kl}}{\Lambda^{2}};
\qquad \bar\Delta_{kl}\ll\Lambda^{2}
\ee
The last inequalities provide the
masses to be essentially less then the cutoff. 

For strong four-fermion coupling
constants $a_{kl} \geq 8\pi^{2}\delta_{kl}$ in the scalar channel
the above interaction induces the dynamical chiral symmetry
breaking. In the language of the theory of critical phenomena it
is equivalent to developing our model around a critical or
scaling point (a critical surface in our case) where the
quantum system undergoes the second-order phase transition. The
parameters $\Delta_{kl}$ just describe the deviation from a
critical point and determine the physical mass of scalar mesons.
The CSB is generated by the dynamic quark mass function corresponding
to nontrivial v.e.v. of scalar fields $\sigma_1,\sigma_2$,
\be
M(\tau) = \sigma_1 f_1(\tau) + \sigma_2 f_1(\tau);\qquad M_0 \equiv
M(0) = 2 \sigma_1.
\ee  
This CSB is characterized by the quark condensate,
\begin{equation}
<\bar q q > \simeq - \frac{N_c\Lambda^2}{8\pi^2}
(\sigma_1 - \sqrt{3} \sigma_2). \label{cond2}
\end{equation}
We remark that this quark condensate in general is not defined solely
by the dynamical mass $M_0$ as in the conventional (one-channel) NJL
model but depends also from the second v.e.v. $\sigma_2$. Only if
the dynamical mass function is strictly constant, 
$\sigma_2 = -\sqrt{3}\sigma_1 $, one reproduces the NJL relation
\be
<\bar q q > \simeq - \frac{N_c\Lambda^2 M_0}{4\pi^2}. \label{cond1}
\ee

When integrating out the quark fields  one comes
to the bosonic effective action where the quadratic parts
in boson field fluctuations are retained only 
to describe the meson mass spectrum,
$$
S_{eff}\simeq\frac{1}{2}\int\frac{d^{4}p}{(2\pi)^{4}}\sum_{k,l=1}^{2}[\sigma_{k}
\Gamma_{kl}^{\sigma\sigma}\sigma_{l}
+
\pi_{k}\Gamma_{kl}^{\pi\pi}\pi_{l}
$$
\be+\rho_{k,\mu}\Gamma_{kl}^{\rho\rho,\mu\nu}\rho_{l,\nu}+a_{k,\mu}
\Gamma_{kl}^{aa,\mu\nu}a_{l,\nu}
+ 2\pi_{k}\Gamma_{kl}^{\pi a,\mu}a_{l,\mu}].
\ee
Herein $\Gamma_{i}$ (except the last one - $\Gamma^{\pi a}$,
which corresponds to the $\pi-a$ mixing ) have the structure:
\be
\hat{\Gamma}=\hat{A}p^{2}+\hat{B},
\ee
where the two symmetric matrices - the kinetic term $\hat{A}$ and
the momentum independent part $\hat{B}$ - have been
introduced. Several comments are in order.
\begin{enumerate}
\item[(i)] We neglect terms $O(\frac{1}{\Lambda^{2}})$ in all
calculations.
\item[(ii)] The large-log approximation
$\ln\frac{\Lambda^{2}}{M_{0}^{2}}\gg 1$ is adopted, where $M_{0}$
denotes the dynamical mass at zero external momentum. It is compatible with
confinement.
\item[(iii)] In order that the physical masses 
were insensitive to $\Lambda$,
 the effective coupling constants 
should be weakly dependent on the cutoff, 
$\Delta_{kl},\, \bar\Delta_{kl}\sim
M_{0}^{2}\ln\frac{\Lambda^{2}}{M_{0}^{2}}$.
\end{enumerate}
The physical mass spectrum can be found from solutions of the
corresponding secular equation,
\be
det(\hat{A}p^{2}+\hat{B})=0;\qquad m_{phys}^{2}=-p_{0}^{2}.
\ee
The consistency of mass equation in the vector and axial-vector
channel imposes the similar scale condition on parameters
$\bar\Delta_{kl}$ as in the scalar and pseudoscalar case:
\be
\bar\Delta_{kl}=O(\ln{\frac{\Lambda^{2}}{M_{0}^{2}}}).
\ee
Let us display the mass-spectrum for ground meson states and their first
excitations. We introduce the notations,
$$
\sigma^2\equiv\sigma_{1}^{2}+
\frac{2\sqrt{3}}{3}\sigma_{1}\sigma_{2}+3\sigma_{2}^{2}>0, \label{sigma}
$$
\be
d\equiv 3\bar\Delta_{11}+2\sqrt{3}\bar\Delta_{12}+\bar\Delta_{22}, \label{de}
\ee
and take into account the consistency inequalities
\be
\Delta_{22}<0,\qquad
\bar\Delta_{22}<0.
\ee
\begin{center}
\underline{Spectrum for scalar and pseudoscalar mesons}
\end{center}
\be
m_{\sigma}=4\sigma_{1}=2M_{0};\qquad
m_{\pi}=0; \label{nambu}
\ee
\be
m_{\pi'}^{2}\simeq-\frac{4}{3}\Delta_{22}+\sigma^2;\qquad
m_{\sigma'}^{2}-m_{\pi'}^{2}\simeq 2\sigma^2>0. \label{sipi}
\ee
\begin{center}
\underline{Spectrum for vector and axial-vector mesons}
\end{center}
\be
m_{\rho}^{2}\simeq
-\frac{\mbox{\rm det}\bar\Delta_{kl}}{2\bar\Delta_{22}
\ln{\frac{\Lambda^{2}}{M_{0}^{2}}}};\qquad
m_{a}^{2}\simeq m_{\rho}^{2}+6M_{0}^{2};\label{arho}
\ee
\be
m_{\rho'}^{2}\simeq-\frac{4}{3}\bar\Delta_{22}-
\frac{d}{6\ln{\frac{\Lambda^{2}}{M_{0}^{2}}}}-m_{\rho}^{2};\qquad
m_{a'}^{2}-m_{\rho'}^{2}\simeq\frac{3}{2}(m_{\sigma'}^{2}-m_{\pi'}^{2})
= 3\sigma^2 >0.
\label{exc}
\ee
The prime labels everywhere the corresponding excited meson state.

Let us comment on these spectra.
\begin{enumerate}
\item[(i)] The formula for $m_{\sigma}$ is a well-known Nambu relation.
\item[(ii)] The expression for $m_{a}$ is also a known result (see, for
example, \cite{[7]}-\cite{[10]}). 
Combining it with the Nambu
relation we arrive to the remarkable equality:
\be
m_{a}^{2}-m_{\rho}^{2}\simeq\frac{3}{2}m_{\sigma}^{2},
\ee
which has the same form as the relation for excited meson states \gl{exc}
for $m_\pi =0$.
\item[(iii)] In both spectra the radial excitations are
logarithmically heavier than the corresponding ground meson states.
\item[(iv)] The value of mass splitting between excited states in 
both spectra is logarithmically suppressed as compared with the mass
values. Therefore the masses of first excited meson
states both in the scalar-pseudoscalar channel and in the
vector-axial-vector one are approximately equal.
\item[(v)] Another prediction is that a mass of $\sigma'$-meson is a
little more than that of $\pi'$-meson. The situation in the
vector-axial-vector spectrum is opposite - the
$a'$-meson is a little heavier than $\rho'$-one.
\item[(vi)] When taking into account the $\pi - a_1$ mixing the 
pion decay constant is determined by the physical meson masses, 
\be
\label{fpim}
f_{\pi}^2\simeq\frac{N_cm_{\sigma}^2m_{\rho}^2}{16\pi^2m_{a_1}^2}
\ln\!\frac{\Lambda^{2}}{M_{0}^{2}}\, ,
\ee
wherefrom it follows that the logarithm of the cutoff 
is uniquely obtained from this relation
in terms of physical parameters.
\end{enumerate}

We identify $\sigma$ with $f_{0}(400-1200)$, $\sigma'$ with
$f_{0}(1370)$, $\pi'$ with $\pi(1300)$, $\rho$ with $\rho(770)$,
$\rho'$ with $\rho(1450)$ and $a$ with $a_{1}(1260)$ \cite{[31]}. The
first excitation of  the $a_{1}$ - the particle $a'$ - is not
found yet. This second axial-vector state, perhaps, could be
found from hadronic $\tau$ decays, although there are the strong
phase space limitations \cite{[20]}.The experimental data give us:
$$
m_{\sigma}=400\div1200 \mbox{ MeV};\qquad
m_{\sigma'}=1200\div1500 \mbox{MeV};\qquad
m_{\pi'}=1300\pm 100 \mbox{ MeV};
$$
\be
m_{\rho}=770\pm 0.8 \mbox{ MeV};\qquad
m_{\rho'}=1465\pm 25 \mbox{ MeV};\qquad
m_{a}=1230\pm 40 \mbox{ MeV}.
\ee
The prediction for mass of $\sigma$-meson is then
\be
m_{\sigma} \simeq 800 \mbox{ MeV}, \label{sig}
\ee
which is close to the averaged experimental value. 
Furthermore we have the following prediction for the mass of
$a'$-particle,
\be
m_{a'}\cong 1465 \div 1850 \mbox{ MeV}.
\ee
The large range for a possible mass of $a'$-meson is accounted for by 
a big experimental uncertainty for the mass of $\sigma'$ and
$\pi'$ mesons. If we accept the averaged values for them, then 
$m_{a'}-m_{\rho'}\approx 30$ MeV.

Let us compare the parameters of the NJL and QNJL models.
First we notice that in both models the Nambu relation \gl{nambu}
together with \gl{arho} leads to the large value of the
dynamical mass $M_0 \simeq 400$ MeV for physical masses of
$\rho$ and $a_1$ mesons. Then the physical value of the pion
decay constant \gl{fpim} $f_\pi \simeq 90$ MeV  can be achieved
for a cutoff $ \Lambda \simeq 1$ GeV. In turn, in the NJL it leads
to a unacceptably high values of the quark condensate \gl{cond1},
$<\bar q q> \simeq - (310)\mbox{MeV}^3$ whereas in the QNJL one can
easily adjust the second v.e.v. to reproduce the condensate 
in agreement with the estimations from Chiral Perturbation Theory
\cite{[32],[34]} (see also,\cite{iof}). In particular, the
value $<\bar q q> \simeq - (250)\mbox{MeV}^3$ corresponds to
$\sigma_1 \simeq -\sqrt{3} \sigma_2$ (see the discussion of this point
 in the next section).
\bigskip

{\large\bf 3. Chiral symmetry restoration constraints }\\

Let us exploit the constraints based on chiral symmetry
restoration at QCD at high energies. Expanding the meson correlators
\gl{cor1} in
powers of $p^{2}$ one arrives to the CSR Sum Rules. In the
scalar-pseudoscalar case \gl{SP} they read:
$$
\sum_{n}Z_{n}^{S}-\sum_{n}Z_{n}^{P}=0;
$$
\be
\sum_{n}Z_{n}^{S}m_{S,n}^{2}-\sum_{n}Z_{n}^{P}m_{P,n}^{2}=\Delta_{SP},
\ee
and in the vector-axial-vector one \gl{VA} one obtains:
$$
\sum_{n}Z_{n}^{V}-\sum_{n}Z_{n}^{A}=4f_{\pi}^{2};
$$
$$
\sum_{n}Z_{n}^{V}m_{V,n}^{2}-\sum_{n}Z_{n}^{A}m_{A,n}^{2}=0,
$$
\be
\sum_{n}Z_{n}^{V}m_{V,n}^{4}-\sum_{n}Z_{n}^{A}m_{A,n}^{4}=\Delta_{VA}.
\ee
The first two relations  are
the famous Weinberg Sum Rules, with $f_{\pi}$  being the
pion decay constant. The residues in resonance pole contributions
in the vector and axial-vector correlators have the structure,
\be
Z_{n}^{(V,A)}=4f_{(V,A),n}^{2}m_{(V,A),n}^{2},
\ee
with $f_{(V,A),n}$ being defined as corresponding
decay constants. 

In the scalar-pseudoscalar case it has been obtained \cite{[24],[46]}
that:
\begin{enumerate}
\item[(i)] the residues in poles are of different order of
magnitude,
\be
\frac{Z_{\sigma,\pi}}{Z_{\sigma',\pi'}}=
O(\frac{1}{\ln\frac{\Lambda^{2}}{M_{0}^{2}}}).
\ee
Moreover, due to $m_{\pi'}^{2}\simeq m_{\sigma'}^{2}$ it happens
to be 
\be
Z_{\pi}\simeq Z_{\sigma};\qquad Z_{\pi'}\simeq Z_{\sigma'} \simeq Z_0 \equiv
\frac{N_c\Lambda^4}{2\pi^2}. \label{znul}
\ee
\item[(ii)] The remarkable relation takes place:
\be
Z_{\sigma}\simeq 4\frac{<\bar{q}q>^{2}}{f_{\pi}^{2}}.
\ee
\item[(iii)] The second CSR Sum Rules constraint results in the
estimation for splitting between the $\sigma'$- and
$\pi'$-meson masses,
\be
Z_0 (m_{\sigma'}^2 -  m_{\pi'}^2) \simeq 24 \pi\alpha_s
< \bar q q >^2. \label{scal}
\ee
If the optimal fit for $f_\pi$ and the
quark condensate is performed then $\sigma_1 \simeq - \sqrt{3}\sigma_2$
(see the end of the previous section) and,
\be
m_{\sigma'}^{2}-m_{\pi'}^{2}\simeq\frac{1}{6}m_{\sigma}^{2}.
\ee
\item[(iv)] The chiral coupling constant $L_{8}$ \cite{[32],[33]} are
calculated:
\be
L_{8}=\frac{f_{\pi}^{4}}{64<\bar{q}q>^{2}}(\frac{Z_{\sigma}}{m_{\sigma}^{2}}
+\frac{Z_{\sigma'}}{m_{\sigma'}^{2}}-\frac{Z_{\pi'}}{m_{\pi'}^{2}})\simeq
\frac{f_{\pi}^{2}}{16m_{\sigma}^{2}}(1-\frac{6\alpha_{s}\pi f_{\pi}^{2}
m_{\sigma}^{2}}{m_{\pi'}^{4}}).
\ee
Its value $L_{8}=(0.9\pm 0.4)\cdot 10^{-3}$ from \cite{[32],[34],[35]}
accepts $m_{\sigma}\simeq 800$ MeV.
\end{enumerate}

In the vector-axial-vector case all residues 
are found to be of the same order of magnitude in contrast to the
scalar-pseudoscalar channel. When 
using the
notation \gl{de}
they read:
$$
Z_{\pi} = 4 f^2_{\pi} \simeq-\frac{N_c\Lambda^4(m_{a_1}^2-m_{\rho}^2)
(6m_{\rho}^2
\ln\!\frac{\Lambda^2}{M_0^2}+ d)}{32\pi^2m_{\rho}^2m_{a_1}^2m_{a_1'}^2
\ln\!\frac{\Lambda^2}{M_0^2}},
$$
$$
Z_{\rho}\simeq\frac{m_{a_1}^2}{m_{a_1}^2-m_{\rho}^2}Z_{\pi};\qquad
Z_{a_1}\simeq\frac{m_{\rho}^2}{m_{a_1}^2-m_{\rho}^2}Z_{\pi};
$$
\be
Z_{\rho'}\simeq\frac{Z_1}{m_{\rho'}^2};\qquad
Z_{a_1'}\simeq\frac{Z_1}{m_{a_1'}^2};
\qquad Z_1\equiv\frac{3N_c\Lambda^4}{16\pi^2}.
\ee
The relation for
$Z_{\pi}$ is considered a
constraint on effective coupling constants of the QNJLM
$\bar{\Delta}_{kl}$ (see \gl{fpim}).
The first and the second Sum Rules are fulfilled identically
in the large-log approach. The third one takes the form:
$$
Z_1(m_{a_1'}^2-m_{\rho'}^2)\simeq16\pi\alpha_s<\!\bar{q}q\!>^2.
\label{3sr}
$$
It is instructive to compare this sum rule
with the similar one, \gl{scal} in the scalar channel. In terms of
\gl{cond2} and \gl{sigma} and taking into account eqs.\gl{sipi},
\gl{exc} one obtains:
$$
\sigma^{2}
\simeq \frac{4N_c\alpha_s }{9\pi}
(\sigma_{1} - \sqrt{3} \sigma_{2})^2, \quad\mbox{vector SR}\quad
\mbox{\it\bf vs.}\quad
\sigma^{2}
\simeq \frac{3N_c\alpha_s }{8\pi}
(\sigma_{1} - \sqrt{3} \sigma_{2})^2, \quad\mbox{scalar SR}.
$$
The minor discrepancy between 4/9 and 3/8 is about 15\% and can be referred to
the quality of two-resonance approximation. Thus the above sum rules
consistently bound v.e.v.'s of scalar fields, i.e. control CSB.

The
structure of $Z_{\rho'}$ and $Z_{a'}$ shows that if $m_{a'}\simeq
m_{\rho'}$ then $Z_{a'}\simeq Z_{\rho'}$ and therefore
$f_{a'}\simeq f_{\rho'}$. As a consequence these residues
approximately cancel each other in Sum Rules and the one-channel
results for $f_{\rho},f_{a}$ holds,
\be
f_{\rho}\simeq\frac{f_{\pi}m_{a}}{m_{\rho}\sqrt{m_{a}^{2}-m_{\rho}^{2}}};\qquad
f_{a}\simeq\frac{f_{\pi}m_{\rho}}{m_{a}\sqrt{m_{a}^{2}-m_{\rho}^{2}}}.
\ee
After evaluating we get $f_{\rho}\approx 0.15$ and $f_{a}\approx
0.06$ to be compared with the experimental values \cite{[36]} from the
decay $\rho^{0}\rightarrow e^{+}e^{-}$, $f_{\rho}=0.20\pm 0.01$,
and from the decay $a_{1}\rightarrow\pi\gamma$, $f_{a}=0.10\pm
0.02$. We have also a
reasonable prediction for the chiral constant $L_{10}$ \cite{[32]}
\be
L_{10}=\frac{1}{4}(\sum_{n}f_{A,n}^{2}-\sum_{n}f_{V,n}^{2})\simeq
\frac{1}{4}(f_{a}^{2}-f_{\rho}^{2})\approx -5.1\cdot 10^{-3},
\ee
which confronts the phenomenologically accepted estimation \cite{[35]}
$L_{10}=-(5.5\pm 0.7)\cdot 10^{-3}$.

\bigskip

{\large\bf 4. Summary}\\

We have shown that the QNJL model truncating
low energy QCD effective action in vector sector can serve to
describe the physics of meson resonances. The matching to
non-perturbative QCD based on chiral symmetry restoration at high
energies improves the predictability of the model. The CSR Sum
Rules are well saturated by four resonances.
Let us summarize the results obtained in the present work.
\begin{enumerate}
\item[(i)] The mass of the second axial-vector particle with $I=1$
is predicted. It is comparable with the mass of the
vector counterpartner: $m_{a'}=1465\div 1850$ MeV and the most
plausible value of the mass difference is
$m_{a'}-m_{\rho'}\approx 30$ MeV.
\item[(ii)] The estimation on the mass of the $\sigma$-meson
does not contradict to existing experimental data \cite{[31]}:
$m_{\sigma} \simeq 800$ MeV.
\item[(iii)] The couplings $f_{\rho},f_{a}$ and the chiral constant
$L_{10}$ are evaluated from CSR Sum Rules 
as matching rules for our model to QCD at
intermediate energies, with the result being $f_{\rho}\approx
0.15,f_{a}\approx0.06,L_{10}\approx -5.1\cdot 10^{-3}$;\\
\end{enumerate}

Finally we would like to mention possible
applications of QNJLM. Firstly, such models are thought of as
relevant for investigations of behaviour of hadron matter at high
temperatures and nuclear densities  in the region near the
restoration of chiral symmetry. One could
expect that for increasing quark densities the mass-splitting is
collapsing and therefore excitations become lighter and more
important in hadron kinetics. Secondly, QNJLM can be used to
describe Higgs particles in extensions of the Standard Model
(SM). The
models of such extension can be found in \cite{[46],[45]}.

\bigskip

{\large\bf Acknowledgements}\\

We express our gratitude to the organizers of the
International Workshop QFTHEP 2000 in Tver' and
especially to Prof.V.I.Savrin and Dr.V.V.Keshek for
hospitality and financial support.
This work is supported by Grant
GRACENAS, by Generalitat de Catalunya,
Grant PIV 1999, and by Grant
RFBR 98-02-18137.

\end{document}